\documentclass[preprint,showpacs,aps]{revtex4}
\usepackage{graphicx}

\begin{document}
\title{%
Experimental realization of a quantum phase transition of 
polaritonic excitations
}
\author{
Kenji Toyoda%
\footnote{e-mail: toyoda@ee.es.osaka-u.ac.jp.}%
, Yuta Matsuno, Atsushi Noguchi, Shinsuke Haze, and Shinji Urabe}
\affiliation{%
Graduate School of Engineering Science, Osaka University,
1-3 Machikaneyama, Toyonaka, Osaka, Japan
}
\date{\today}

\begin{abstract}
%
%
%
%
We report an experimental realization of the Jaynes-Cummings-Hubbard
(JCH) model using the internal and radial phonon states of two trapped
ions.  An adiabatic transfer corresponding to a quantum phase transition
from a localized insulator ground state to a delocalized superfluid (SF)
ground state is demonstrated.  The SF phase of polaritonic excitations
characteristic of the interconnected Jaynes-Cummings (JC) system is
experimentally explored, where a polaritonic excitation refers to a
combination of an atomic excitation and a phonon interchanged via a JC
coupling.
\end{abstract}


\maketitle

The Jaynes-Cummings (JC) model
\cite{Jaynes1963,Meekhof1996}
describing the interaction between
a quantized optical mode and a two-level atom
is one of the simplest and most important models of light-matter 
interactions.
An interconnected array of multiple JC systems
has recently been attracting interest, and the model describing it is referred to as
the Jaynes-Cummings-Hubbard (JCH) model
\cite{Greentree2006,Hartmann2006,Angelakis2007,Rossini2007,Irish2008,Makin2008,Ivanov2009};
an experimental realization of this model has remained to be done.
The JCH model was originally proposed for an array of coupled optical
cavities, each containing a two-level atom, and is expected to exhibit 
properties peculiar to strongly correlated systems 
\cite{Hubbard1963,Vanderzant1992,Greiner2002}.

In the JCH model for an array of coupled optical cavities,
photons naturally hop between neighboring cavities,
whereas the photon-photon interaction arises from a photon blockade
\cite{Rebic2002},
which impedes other photons from entering an occupied cavity.

The JCH model has certain similarities to the Bose-Hubbard (BH) model
\cite{Vanderzant1992,Greiner2002}. 
It approaches the pure bosonic case in
the large detuning and the large photon number limits \cite{Greentree2006}.
In contrast, in the limit of small detuning and small
phonon numbers, the coefficient for the on-site repulsion becomes
dependent on the photon number. In addition, the 
conserved particles (polaritons or dressed atoms)
transform into various kinds of excitations (atomic excitations, photons or
polaritons) depending on the Rabi frequency and detuning.
As a result, a JCH system has a richer phase structure compared with 
a BH system. Both photons and polaritons can show superfluidity, while 
insulator phases can be formed with both atoms and polaritons 
\cite{Rossini2007,Irish2008}.

Recent advances in the ability to manipulate quantum systems
have made it possible to simulate a quantum system using another
controllable system (analog quantum computation)
\cite{Feynman1982}.
Trapped 
ions offer high controllability and individual access, and hence are
suited for such applications
\cite{Johanning2009,Blatt2012}. 
Simulations of systems including spin
systems and relativistic electrons have been reported
\cite{Friedenauer2008,Gerritsma2010,Kim2010,Lanyon2011,Islam2011}.
Simulations of Hubbard models have also been proposed for trapped ions
\cite{Porras2004,Ivanov2009}, however
an experimental demonstration has remained to be done.

The phonons in the radial (or transverse) direction of a linear ionic
chain, which have been used to mediate spin-spin interactions%
\cite{Kim2010,Islam2011},
can also be used to simulate systems of Bosonic particles
under certain conditions \cite{Porras2004}.
In contrast to the axial motion of ions in a linear chain,
which is described by collective modes that span over the
whole ionic chain,
radial phonons under a sufficiently tight radial confinement 
are essentially 'local phonons' (phonons of local harmonic oscillations) 
undergoing hopping from site to site
with a rate slower than the local harmonic-oscillation frequencies. 
We recently observed this hopping of radial phonons using two trapped
ions\cite{Haze2012}.
Ivanov {\it et al.}\cite{Ivanov2009} proposed to
use a JC coupling arising from optical excitation of the radial red sideband
transition of a linear ionic chain
to induce an effective phonon-phonon coupling, thereby
realizing the JCH model.
In this letter, we report an experimental realization of the 
JCH model and observation of 
a quantum phase transition based on Ivanov {\it et al.}\cite{Ivanov2009}\,\,\,
In this case the conserved particles
are not merely phonons but
composite particles each of which is a linear combination 
of a phonon and an atomic excitation.

\begin{figure}[]
\includegraphics[width=8cm]{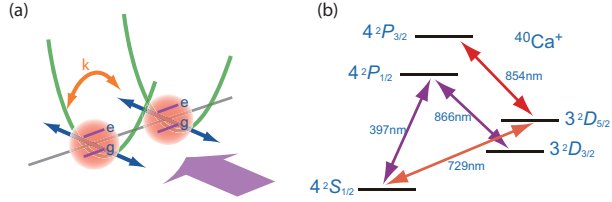} 
\caption{ 
(color online).
(a) Conceptual schematic of the JCH system with two ions.
Two ions are illuminated with an excitation laser nearly resonant to
the red-sideband transition with which
the JC coupling arises and dressed atoms or polaritons
are formed.
Inter-site hopping ($\kappa$) is naturally incorporated from Coulomb couplings
and along with effective on-site repulsion between polaritons due to the JC
 coupling,
a Hubbard-type model is formed.
(b) Energy levels of $^{40}$Ca$^+$ relevant to motional cooling and induction of the JC coupling.
}
\label{fig:jch_conceptualschematic}
\end{figure}

The conceptual schematic of the JCH system using trapped ions
is shown in Fig.\ \ref{fig:jch_conceptualschematic}(a).
It is assumed that two ions with internal states
$\{\left|g_j\right\rangle,\left|e_j\right\rangle\}$ and
a resonance frequency $\omega_0$
are held in a linear Paul trap.
Each of the ions undergoes harmonic motion in a radial direction
(referred to as the $x$ direction). 
Both ions are equally illuminated with a laser 
of frequency $\omega_L$ and detuning
$\delta=\omega_L-\omega_0$, which is 
nearly resonant with the radial red-sideband transitions.
Then the system is approximately 
governed by the following Hamiltonian
(a JCH Hamiltonian)
\cite{Ivanov2009}:
\begin{eqnarray}
H&=&\hbar\Delta \sum_{j=1,2}
\left|e_j\right\rangle\left\langle{}e_j\right|
+\hbar{}g \sum_{j=1,2} \left(
\hat{a}_j^\dagger \hat{\sigma}^-_j + \hat{a}_j \hat{\sigma}^+_j 
\right)\nonumber
\\
&&+\frac{\kappa}{2}\left( \hat{a}_1^\dagger \hat{a}_2 + \hat{a}_2^\dagger \hat{a}_1 \right).
\label{eqn:JCH}
\end{eqnarray}
Here
$\Delta\equiv-\delta-\omega_x'=-[\omega_L-(\omega_0-\omega_x')]$
is the detuning from the radial red-sideband transition,
where
$\omega_x'\equiv\omega_x+\Delta\omega_x$ is the 
oscillation frequency for the radial $x$ direction,
in which $\omega_x$ is the oscillation frequency of a single ion in the radial $x$ direction 
and
$\Delta\omega_x=-\omega_z^2/4\omega_x=-e^2/8\pi\epsilon_0 d_0^3 m
\omega_x$ is the correction due to the Coulomb interaction %
($d_0$ is the inter-ion distance and $m$ is the mass of one ion).
$g\equiv{\eta\Omega_0}/{2}$ is the coupling coefficient for the 
red sideband transition (JC coupling)
where 
$\eta$ is the Lamb-Dicke factor and 
$\Omega_0$ is the on-resonance Rabi frequency.
$\hat{a}_j^\dagger$ and $\hat{a}_j$ are the creation and annihilation
operators of phonons in the radial $x$-direction of the $j$-th ion,
whose Hilbert space is spanned 
by Fock-state basis  $\left|n\right\rangle_j$ ($n=0,1,2,...$).
$\hat{\sigma}^+_j \equiv \left|e_j\right\rangle\left\langle{}g_j\right|$ and
$\hat{\sigma}^-_j \equiv \left|g_j\right\rangle\left\langle{}e_j\right|$
are the raising and lowering operators for the internal states.
$\kappa=\omega_z^2/2\omega_x=e^2/4\pi\epsilon_0 d_0^3 m \omega_x$ is the 
hopping rate for the radial $x$ direction.
%


A JCH system is expected to show quantum phase transitions
between superfluid and insulator phases of polaritons
\cite{Angelakis2007}.
Here a 'superfluid' is a system
that has delocalized excitations and in which 
there is a correlation between mechanical variables at different sites.
On the other hand, an 'insulator' is a system 
that has localized excitations. 

As an order parameter characterizing the quantum phase transition,
the variance of the total excitation number per site, 
$\Delta \hat{N}_j^2 \equiv
\langle\hat{N}_j^2\rangle-\langle\hat{N}_j\rangle^2$, 
where 
$\hat{N}_j=\hat{a}_j^\dagger{}\hat{a}_j+\left|e_j\right\rangle\left\langle{}e_j\right|$,
can be used \cite{Angelakis2007}.
The expectation value of the annihilation operator
which is usually used in the mean-field limit
cannot be used as the order parameter,
since it is always zero for a closed system  
with no particle exchange
with the outside\cite{Angelakis2007,Irish2008}.
In addition, the atomic excitation number variance 
$\Delta \hat{N}_{a,j}^2 \equiv
\langle\hat{N}_{a,j}^2\rangle-\langle\hat{N}_{a,j}\rangle^2$, 
where
$\hat{N}_{a,j}=\left|e_j\right\rangle\left\langle{}e_j\right|$,
is also used for judging
the existence of polaritons.



The experimental setup used is similar to that described in
\cite{Haze2012} and a brief description is given here.  Two
$^{40}$Ca$^+$ ions are trapped in vacuum ($5\times10^{-9}$ Pa) using a
linear Paul trap.  
A RF voltage of 25 MHz is applied to generate
the radial confinements and DC electrodes provide an axial confinement.
The secular frequencies for the three trap axes
are $(\omega_x,\omega_y,\omega_z)/2 \pi=(2.1,1.7,0.17)$ MHz.
The inter-ion distance in the axial direction $d_0$
is 18--20 $\mu$m and correspondingly, the hopping rate
$\kappa/2\pi$ is 5--7 kHz.

The energy levels relevant to motional cooling
and induction of the JC coupling are shown in 
Fig.\ \ref{fig:jch_conceptualschematic}(b).
The motion in the radial directions is cooled by
Doppler cooling using $S_{1/2}$--$P_{1/2}$ (397 nm) and
$D_{3/2}$--$P_{1/2}$(866 nm) 
and sideband cooling using $S_{1/2}(m_J=-1/2)$--$D_{5/2}(m_J=-5/2)$ (729
nm) and $D_{5/2}$--$P_{3/2}$(854 nm).
There is two collective modes in the $x$ direction of two ions,
namely the center-of-mass (COM; in-phase) mode and the rocking
(out-of-phase) mode, just as in the case of the axial motion
\cite{James1998}.
The average quantum numbers after the sideband cooling are
$(\bar{n}_{x,{\rm COM}},\bar{n}_{x,{\rm Rock}},\bar{n}_{y,{\rm COM}},\bar{n}_{y,{\rm Rock}})=(0.04\pm0.04,0.03\pm0.03,0.57\pm0.11,0.08\pm0.04)$.
The axial motion is cooled only by Doppler cooling.
The ions are intermittently
optically pumped to $S_{1/2}(m_J=-1/2)$ by using a 397-nm beam
with the $\sigma^-$ polarization
during and after the sideband cooling.
The excitation beam at 729 nm for the $S_{1/2}$--$D_{5/2}$ transition,
which is used to induce the JC coupling and other
operations, is 
oriented at 45$^\circ$,45$^\circ$, and 90$^\circ$ relative to
the $x$, $y$, and $z$ directions, respectively.
This direction is chosen to couple the beam only to the radial directions
and to ignore the axial directions, whose secular frequency is relatively
small and hence less advantageous in sideband cooling because of the
large average quantum number after Doppler cooling.
Equal illumination of the two ions with this beam is carefully optimized by 
adjusting the beam position so that
the intensity difference between the two ions becomes less than 5\%.

The internal state of the ions is determined by illuminating them with
lasers at 397 nm ($S_{1/2}$--$P_{1/2}$ transition) 
and 866 nm ($D_{3/2}$--$P_{1/2}$ transition) and by
detecting fluorescence photons with a photomultiplier or an intensified
charge-coupled-device (ICCD) camera, with detection times of
8 ms and 80 ms, respectively.
Individual detection of fluorescence from each ion is possible with the ICCD camera.
Due to unequal illumination intensity of the two
ions with the 397 nm laser,
individual detection is possible also when using the photomultiplier.

\begin{figure}[]
\includegraphics[width=8cm]{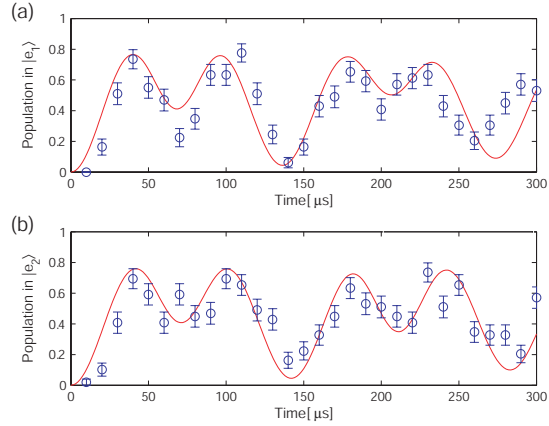} 
\caption{ 
(color online).
Measured and simulated quantum dynamics of the JCH
 system with two ions.
Populations of the excited state of one ion and the other 
are plotted in (a) and (b), respectively, with circles.
Each point obtained is the average of 50 experiments.
Curves are numerically simulated results,
which are multiplied by 0.8 to consider
population quenching from $D_{5/2}$ to $S_{1/2}$ in the relatively
long detection time of 80 ms, due to 
a stray intensity from the 854-nm beam.
}
\label{fig:jch_dynamics}
\end{figure}

First, the dynamics of the JCH system with two ions is observed. 
The total dynamics of the two-ion JCH system 
arises from the JC coupling in individual atoms
excited by the excitation laser and inter-site hopping\cite{Haze2012}.
When the hopping rate $\kappa$ is much smaller than the JC
coupling coefficient $g$, a simple sinusoidal oscillation similar
to Rabi dynamics caused by
the sideband excitation of the local radial oscillation modes
is expected, while for non-negligible values of $\kappa$, an interference
between Rabi and hopping dynamics is expected.
Figure \ref{fig:jch_dynamics} shows the result of the observation of the JCH dynamics
for two ions (the circles), where
the population of the internal state of each ion is plotted.
The system is initially prepared in the
$\left|g_1\right\rangle \left|g_2\right\rangle
\left|0\right\rangle_1 \left|0\right\rangle_2$
state with sideband cooling and optical pumping.
The excitation laser is tuned to the resonance of the 
blue-sideband transition of the radial $x$ mode.
This gives rise to an anti-Jaynes-Cummings coupling
\cite{Meekhof1996}, 
which is formally equivalent
to a JC coupling when the internal states
$\{\left|g_j\right\rangle,\left|e_j\right\rangle\}$ are
interchanged.
The red curves show numerically simulated dynamics
for the hopping rate $\kappa/2\pi$ of 5.4 kHz, 
the JC coupling coefficient $2g/2\pi$ of 12.0 kHz, and the 
coherence relaxation due to laser frequency fluctuations of 200 Hz.
Although the dynamics is periodical, it is greatly modified from
a simple
sinusoidal oscillation due to the effect of the inter-site hopping term. 
The two ions show almost the same dynamics as expected from equal
illumination.

\begin{figure}[]
\includegraphics[width=8cm]{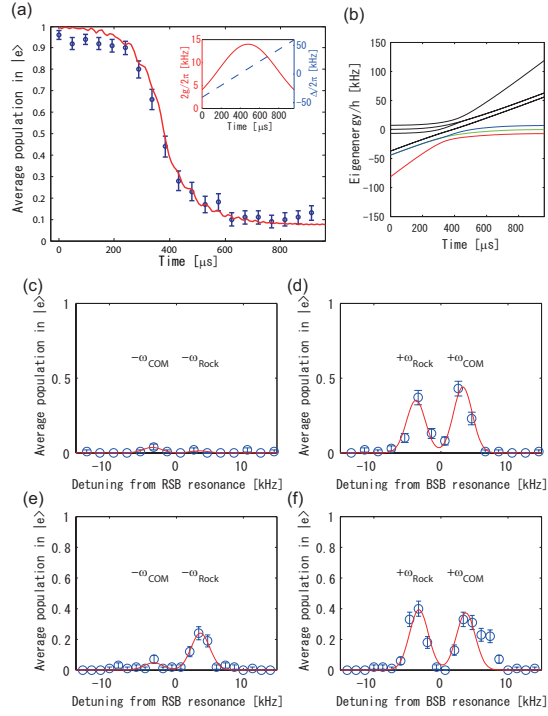}
\caption{
(color online).
(a) Variation of the average internal-state populations
during the adiabatic transfer.
Each point obtained is the average of 50 experiments.
The inset shows the time dependence of 
the JC coupling coefficient $2g/2\pi$
(the solid curve with the vertical axis on the left) and the detuning $\Delta/2\pi$ (the
 dashed curve with the vertical axis on the right).
(b) Time dependence of the eigenenergies
obtained by diagonalyzing the instantaneous Hamiltonians based on the 
pulse parameters used in the experiment.
The lowest three curves corresponds,
from the lowest to the third lowest,
to
$\left|\psi_{\rm atI}\right\rangle\rightarrow
\left|\psi_{\rm phSF}\right\rangle$,
$\frac{1}{\sqrt{2}}
 (\left|g_1\right\rangle \left|e_2\right\rangle+
 \left|e_1\right\rangle \left|g_2\right\rangle)
\otimes
\hat{a}_r^{\dagger}
\left|0\right\rangle_1\left|0\right\rangle_2
\rightarrow
\left|g_1\right\rangle \left|g_2\right\rangle 
\otimes
\hat{a}_c^{\dagger} \hat{a}_r^{\dagger}
\left|0\right\rangle_1\left|0\right\rangle_2$
and
$\frac{1}{\sqrt{2}}
 (\left|g_1\right\rangle \left|e_2\right\rangle-
 \left|e_1\right\rangle \left|g_2\right\rangle)
\otimes
\hat{a}_r^{\dagger}
\left|0\right\rangle_1\left|0\right\rangle_2
\rightarrow
 \left|g_1\right\rangle \left|g_2\right\rangle \otimes
\frac{1}{\sqrt{2}}
\hat{a}_c^{\dagger{}2}
\left|0\right\rangle_1\left|0\right\rangle_2$, 
respectively.
%
(c)-(d) Measurement of the average phonon numbers of collective modes
before the adiabatic transfer.
Results of spectroscopy over radial red and blue
sideband transitions before the adiabatic transfer are shown
in (c) and (d), respectively.
Each point obtained is the average of 50 experiments,
and the red curves are the results of fitting with multiple Gaussians.
The label $-\omega_{\rm COM}$ ($+\omega_{\rm COM}$ ) indicates
the red (blue) sideband resonance of the center-of-mass (COM) mode,
and 
$-\omega_{\rm Rock}$ ($+\omega_{\rm Rock}$ )
the red (blue) sideband resonance of the rocking mode.
(e)-(f) Measurement of the average phonon numbers of collective modes
after the adiabatic transfer, in a similar way to (c)-(d).
Results of spectroscopy over radial red and blue
sideband transitions after the adiabatic transfer
are shown in (e) and (f), respectively.
}
\label{fig:jch_adiabatic}
\end{figure}

As a demonstration of a quantum phase transition,
an adiabatic transfer from an insulator ground state to a SF ground state is observed in the average
excited-state population of two ions 
[Fig.\ \ref{fig:jch_adiabatic} (a)].

The transfer process starts from a point where $-\Delta/g$ is large, where
the ground state is approximately
$\left|\psi_{\rm atI}\right\rangle\equiv
\left|e_1\right\rangle \left|e_2\right\rangle
\left|0\right\rangle_1 \left|0\right\rangle_2$
(the atomic insulator phase).
Then $\Delta/g$ increases, exceeds 
zero and becomes a large positive value,
where the ground state is approximately
$\left|\psi_{\rm phSF}\right\rangle \equiv
 \left|g_1\right\rangle \left|g_2\right\rangle \otimes
(\frac{1}{\sqrt{2}} \left|1\right\rangle_1 \left|1\right\rangle_2
-\frac{1}{2}        \left|2\right\rangle_1 \left|0\right\rangle_2
-\frac{1}{2}        \left|0\right\rangle_1 \left|2\right\rangle_2)=
 \left|g_1\right\rangle \left|g_2\right\rangle \otimes
\frac{1}{\sqrt{2}}\hat{a}_r^{\dagger{}2}\left|0\right\rangle_1
\left|0\right\rangle_2$.
Here,
$\hat{a}_r^{\dagger{}}=\frac{1}{\sqrt{2}}(\hat{a}_1^\dagger-\hat{a}_2^\dagger)$
is the rocking-mode creation operator.
This phase is the phonon SF phase.
In the intermediate region around $\Delta\sim0$,
the system is in the polaritonic SF state.
This polaritonic SF state is approximated as
$\frac{1}{\sqrt{3}} \left|\psi_{\rm phSF}\right\rangle
+\frac{1}{\sqrt{6}} \left|\psi_{\rm atI}\right\rangle
+\frac{1}{2\sqrt{2}}
(\left|e_1\right\rangle \left|g_2\right\rangle 
 \left|1\right\rangle_1 \left|0\right\rangle_2
+\left|g_1\right\rangle \left|e_2\right\rangle 
 \left|0\right\rangle_1 \left|1\right\rangle_2
-\left|e_1\right\rangle \left|g_2\right\rangle 
 \left|0\right\rangle_1 \left|1\right\rangle_2
-\left|g_1\right\rangle \left|e_2\right\rangle 
 \left|1\right\rangle_1 \left|0\right\rangle_2)
$\cite{Irish2008}.

In the experiment, 
$\left|\psi_{\rm atI}\right\rangle$
is prepared with cooling, optical pumping and applying a carrier $\pi$ pulse.
Then the adiabatic
transfer is realized by shining the excitation laser
and sweeping its detuning $\Delta$ 
over the red-sideband resonance from negative to positive values.
The amplitude of the beam is also modulated in a Gaussian 
shape to ensure that $|\Delta|/g$ is large at the beginning and end
of the pulse so that
the overlap of the initial (final) state 
and  $\left|\psi_{\rm atI}\right\rangle$ 
    ($\left|\psi_{\rm phSF}\right\rangle$)
is optimized.
The explicit values of the parameters are as follows.
$\Delta/2\pi$ is linearly swept from -41 kHz to 59 kHz
in 960 $\mu$s, and 
the JC coupling coefficient $2g/2\pi$
is varied from 0.29$\times$14 kHz to 14 kHz and back to 0.29$\times$14 kHz
in a Gaussian shape over the same period
[see the inset of Fig.\ \ref{fig:jch_adiabatic} (a)].
The red curve in Fig.\ \ref{fig:jch_adiabatic} (a) is a numerically simulated result.

The initial population in Fig.\ \ref{fig:jch_adiabatic} (a)
is $\sim$5 \% smaller than what is expected for
$\left|\psi_{\rm atI}\right\rangle$.
This is the result of infidelity in
the carrier $\pi$ pulse used to prepare $\left|\psi_{\rm atI}\right\rangle$,
which is presumably due to jitter in the excitation beam.
The final population is floating from zero by 
$\sim$10 \%.
In addition to the imperfect initialization mentioned above,
this is due also to infidelity in the adiabatic transfer process itself,
which we speculate is due mainly to the effect of laser frequency
fluctuations.
We previously analyzed the effects of laser frequency fluctuations and 
adiabaticity in the rapid adiabatic passage on the
sideband transitions (Fig.\ 4 of \cite{Watanabe2011}; 
although this analysis was done for a single ion,
the overall qualitative and quantitative behavior should be similar).
We have confirmed in a numerical simulation
that the population
go to near zero with less than 1 \% error under the assumption of no
laser 
frequency fluctuation.
Hence we speculate that the effect of diabatic transitions is
limited to below 1 \%.

We also analyzed the adiabaticity during the transfer process using
the theory of adiabatic variations of Hamiltonians\cite{Messiah1961}.
Fig.\ \ref{fig:jch_adiabatic} (b) shows the time-dependent eigenenergies
obtained by diagonalyzing the instantaneous Hamiltonians based on the 
pulse parameters used in the experiment.
From these eigenenergies and eigenvectors, the probability of
diabatic transitions is estimated in a similar way as in \cite{Watanabe2011}.
The largest leakage from the ground state is 
the one towards the third lowest level, and its probability is at most 2 \%. 
This is consistent with the numerical result given above.

The effect of the adiabatic transfer process on the phonon states
is also examined.
Figure \ref{fig:jch_adiabatic} (c)-(f) shows the result of phonon-number
measurements.
Figure \ref{fig:jch_adiabatic} (c) and (d)
show the results of spectroscopy over the radial red- and blue-sideband transitions,
respectively, at the beginning of the adiabatic transfer process, 
and Figure \ref{fig:jch_adiabatic} (e) and (f) 
show the corresponding results at the end of the process.
From these results,
the average phonon numbers for the COM and rocking modes 
at the beginning and end are estimated to be
$(\bar{n}_{\rm COM},\bar{n}_{\rm Rock})=(0.09\pm0.04,0.04\pm0.05)$
and $(\bar{n}_{\rm COM},\bar{n}_{\rm Rock})=(0.15\pm0.11,1.58\pm0.60)$, respectively.
At the beginning, both of the phonon modes are almost in 
the ground states,
while at the end, a number of rocking-mode phonon quanta close to two
is realized and the COM mode is almost intact.
%
The above results support the occurrence of
a quantum phase transition from
the atomic insulator ground state
$\left|\psi_{\rm atI}\right\rangle=
\left|e_1\right\rangle \left|e_2\right\rangle
\left|0\right\rangle_1 \left|0\right\rangle_2$
to the phonon SF ground state
$\left|\psi_{\rm phSF}\right\rangle
 \equiv
 \left|g_1\right\rangle \left|g_2\right\rangle \otimes
\frac{1}{\sqrt{2}}\hat{a}_r^{\dagger{}2}\left|0\right\rangle_1
\left|0\right\rangle_2$.

\begin{figure}[]
\includegraphics[width=8cm]{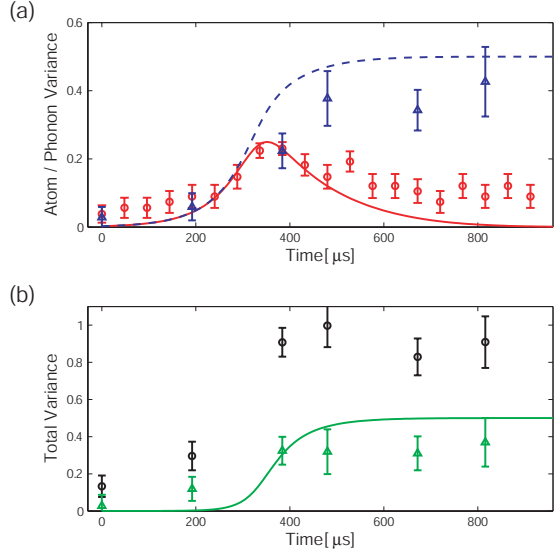} 
\caption{
(color online).
Estimated experimental and
calculated values for the excitation number variances (atomic, phonon and total) 
during the adiabatic transfer.
The points are experimental values and the curves are
calculated from exact ground states
against the actual time dependence of the experimental parameters.
(a) Values for the atomic excitation number
 variance $\Delta \hat{N}_{a,1}^2$ 
(the circles and the solid curve) and
the phonon number variance $\Delta \hat{N}_{p,1}^2$ 
(the triangles and the dashed curve).
Each point obtained is the average of 50 experiments.
(b) Values for the total excitation number variance 
$\Delta \hat{N}_{1}^2$.
Instead of directly estimating the experimental values for this quantity, 
the upper and lower bounds are estimated along with their errors,
and shown as the circles and triangles
(see text for details).
}
\label{fig:jch_variances_exp}
\end{figure}

The transfer process is further analyzed
by estimating the excitation number variances 
(atomic, phonon and total) introduced above.
The red circles in Fig.\ \ref{fig:jch_variances_exp} (a) show 
the atomic excitation number variance 
$\Delta\hat{N}_{a,1}^2$
estimated from atomic populations measured with the photomultiplier tube.
The peak at the center indicates the presence of polaritons.
The numerically simulated results are also shown as the red solid curve.
The cause for the discrepancy between the experimental and calculated
values
is expected to be similar to that discussed in relation to Fig.\
\ref{fig:jch_adiabatic} (a).
The blue triangles in Fig.\ \ref{fig:jch_variances_exp} (a) show
the values of the phonon number variance 
$\Delta\hat{N}_{p,1}^2$
with $\hat{N}_{p,j}=\hat{a}_j^\dagger{}\hat{a}_j$,
which are obtained by measuring the average phonon numbers in the same way as in
Fig.\ \ref{fig:jch_adiabatic} (c)-(f), and estimating
the variance according to 
equation (3) and (4) of the Supplemental material.
This supports the argument that the phonon SF ground state is realized at the end of the
adiabatic transfer.

Estimation of the total excitation number variance
$\Delta\hat{N}_1^2$ requires
simultaneous measurements of internal and phonon states,
which are relatively difficult to perform since phonon states should be
once mapped to internal states to be read.
We avoid such measurements here and instead estimate $\Delta\hat{N}_1^2$
from the known quantities 
$\Delta\hat{N}_{a,1}^2$ and $\Delta\hat{N}_{p,1}^2$.
It should be noted that in this case $\Delta\hat{N}_1^2$ can only be estimated
as intervals with upper and lower bounds.
The details of the derivation of inequalities for estimating
the upper and lower bounds of $\Delta\hat{N}_1^2$ are
given in the Supplemental material.

Figure \ref{fig:jch_variances_exp} (b) 
is the total excitation number variance
$\Delta\hat{N}_1^2$.
The values (upper and lower bounds) are obtained using 
equation (5) and (6) of the Supplemental material
along with the results in 
Fig.\ \ref{fig:jch_variances_exp} (a).
The expected qualitative behavior, including the onset of a phase
transition
(near 400 $\mu$s), an example of which is seen in Fig.\ 2 of \cite{Angelakis2007}, 
is reproduced in these results.

In summary, 
we have observed dynamics and adiabatic transfer between ground states of
a JCH system with two ions. 
Scaling up the JCH system described in this article to include a
larger numbers of sites necessitates certain points to be overcome.
When the number of ions in the linear chain $N_{\rm ions}$ is increased, 
the spacing at the center $d_0$ 
decreases in proportion to $(N_{\rm ions})^{-0.559}$ \cite{James1998}
and hence $\kappa$ increases in proportion to $(N_{\rm ions})^{1.677}$.
On the other hand, it is desired to keep $\kappa/g$ at
moderate values so that 
the rich phase diagram of the JCH system should be explored
as widely as possible.
This demand may be fulfilled by tightening the radial confinement
(note that $\kappa\propto\omega_x^{-1}$) or
by using an array of independent traps, for which inter-ion distances
and magnitudes of confinement can be chosen independently.

\section*{Acknowledgments}
This work was supported by the Ministry of Education, Culture, Sports,
Science and Technology (MEXT) of Japan Kakenhi "Quantum Cybernetics"
project, and by 
the Japan Society for the Promotion of Science (JSPS) through its
Funding Program for World-Leading Innovative R\&D on Science and
Technology (FIRST Program).

\newpage

\newpage
\newpage

%
\end{document}


\title{%
Experimental realization of a quantum phase transition of 
polaritonic excitations
}
\author{
Kenji Toyoda%
, Yuta Matsuno, Atsushi Noguchi, Shinsuke Haze, and Shinji Urabe}
\affiliation{%
Graduate School of Engineering Science, Osaka University,
1-3 Machikaneyama, Toyonaka, Osaka, Japan
}
\date{\today}


\maketitle

\section*{Estimation of the phonon number variance
 and the total
excitation number variances}
In order to determine the total excitation number variance
$\Delta\hat{N}_1^2$,
one has to know the number-state distribution of radial phonons 
per site.
This is not straightforward under the present experimental conditions
since the hopping rate, 
which is currently not dynamically tunable
and is typically set to be 5--7 kHz,
is comparable to the available maximum of the sideband Rabi frequency
($\sim$20 kHz);
hence, during illumination of sideband Rabi pulses
for phonon-number analysis,
the occurrence of hopping is non-negligible.
Instead, we perform measurements of radial collective modes 
to infer populations of local phonons.

When it is assumed that
the COM mode is almost in the ground state, which is the case in the 
present experiment (see Fig.\ 3b-e in the main text),
the expectation values of the phonon number
operator and its square
can be approximated as follows:
%
%
\begin{eqnarray}
 \langle \hat{N}_{p,1} \rangle
&=&\frac{1}{2} \langle 
(\hat{a}_c^\dagger + \hat{a}_r^\dagger) 
(\hat{a}_c         + \hat{a}_r        ) 
\rangle \nonumber \\
&\sim&\frac{1}{2} \langle 
 \hat{a}_r^\dagger \hat{a}_r
\rangle \nonumber \\
&=&
 \frac{1}{2} \langle \hat{N}_{p,r}   \rangle,
\end{eqnarray}
\begin{eqnarray}
 \langle \hat{N}_{p,1}^2 \rangle
&=&\frac{1}{4} \left\langle 
(\hat{a}_c^\dagger + \hat{a}_r^\dagger) 
(\hat{a}_c         + \hat{a}_r        ) 
(\hat{a}_c^\dagger + \hat{a}_r^\dagger) 
(\hat{a}_c         + \hat{a}_r        ) 
\right\rangle \nonumber \\
&\sim&\frac{1}{4} \langle 
\hat{a}_r^\dagger \hat{a}_r \hat{a}_r^\dagger \hat{a}_r
+\hat{a}_r^\dagger \hat{a}_c \hat{a}_c^\dagger \hat{a}_r
\rangle  \nonumber \\
%
&=&\frac{1}{4} \langle 
 \hat{a}_r^\dagger \hat{a}_r \hat{a}_r^\dagger \hat{a}_r
+\hat{a}_r^\dagger \hat{a}_r \hat{a}_c^\dagger \hat{a}_c  
+\hat{a}_r^\dagger \hat{a}_r
\rangle  \nonumber \\
&\sim&
 \frac{1}{4} \langle \hat{N}_{p,r}   \rangle
+\frac{1}{4} \langle \hat{N}_{p,r}^2 \rangle.
\end{eqnarray}
Here, 
$\hat{a}_c^{\dagger{}}=\frac{1}{\sqrt{2}}(\hat{a}_1^\dagger+\hat{a}_2^\dagger)$
and
$\hat{a}_c            =\frac{1}{\sqrt{2}}(\hat{a}_1        +\hat{a}_2)$
are the COM-mode creation and annihilation operators, 
respectively, and  $\hat{a}_r^{\dagger{}}=\frac{1}{\sqrt{2}}(\hat{a}_1^\dagger-\hat{a}_2^\dagger)$
and $\hat{a}_r            =\frac{1}{\sqrt{2}}(\hat{a}_1        -\hat{a}_2)$ are the corresponding rocking-mode operators.
$\hat{N}_{p,r}\equiv\hat{a}_r^\dagger \hat{a}_r$ is the phonon number
operator
for the rocking mode.
Using the above quantities, the phonon number variance can be approximated as follow:
\begin{eqnarray}
 \Delta \hat{N}_{p,1}^2
&\equiv&
 \langle \hat{N}_{p,1}^2 \rangle
-\langle \hat{N}_{p,1} \rangle^2 \nonumber \\
&\sim&
 \frac{1}{4} \langle \hat{N}_{p,r}   \rangle
+\frac{1}{4} \langle \hat{N}_{p,r}^2 \rangle
-\frac{1}{4} \langle \hat{N}_{p,r}   \rangle^2 \nonumber \\
&=&
 \frac{1}{4} \langle \hat{N}_{p,r}   \rangle
+\frac{1}{4} \Delta \hat{N}_{p,r}^2
\end{eqnarray}

To estimate the rocking-mode phonon number variance $\Delta \hat{N}_{p,r}^2$,
one needs to know the population
of rocking-mode phonon Fock states.
For the present experimental conditions, this is difficult
since,
in order to analyze the rocking-mode phonon-number distribution
by exciting the sideband Rabi oscillation,
the sideband Rabi frequency should be much smaller than the hopping rate
(5--7 kHz) to resolve the rocking-mode sideband transition,
and in such a case the sideband Rabi frequency becomes comparable to the dephasing caused by the fluctuations in laser frequency 
(200--300 Hz). 

We use the fact the total number of excitations in the ion chain, $N$ (in this case two),
is conserved for the JCH system to infer $\Delta \hat{N}_{p,r}^2$
from the atomic populations.
When the COM-mode population is neglected and hence
$\hat{N}_{p,r}+\hat{N}_{a,1}+\hat{N}_{a,2}\sim{}N$ is satisfied,
$\Delta \hat{N}_{p,r}^2$ can be expressed using
the atomic excitation number variance, as follows,
%
\begin{eqnarray}
 \Delta \hat{N}_{p,r}^2
&\equiv&
 \langle \hat{N}_{p,r}^2 \rangle
-\langle \hat{N}_{p,r} \rangle^2 \nonumber \\
&=&
\langle
N^2-2 N (\hat{N}_{a,1}+\hat{N}_{a,2}) + (\hat{N}_{a,1}+\hat{N}_{a,2})^2
\rangle \nonumber \\
&&-
\langle
N-(\hat{N}_{a,1}+\hat{N}_{a,2})
\rangle^2 \nonumber \\
&=&
\langle
(\hat{N}_{a,1}+\hat{N}_{a,2})^2
\rangle
-
\langle
\hat{N}_{a,1}+\hat{N}_{a,2}
\rangle^2 \nonumber \\
&=&
 \Delta \hat{N}_{a,1}
+\Delta \hat{N}_{a,2}
+2 \mbox{Cov}(\hat{N}_{a,1},\hat{N}_{a,2})
\end{eqnarray}
where $\mbox{Cov}(\hat{N}_{a,1},\hat{N}_{a,2})\equiv
 \langle \hat{N}_{a,1} \hat{N}_{a,2} \rangle
-\langle \hat{N}_{a,1} \rangle \langle \hat{N}_{a,2} \rangle$
is the covariance (cross correlation) between the two operators.
The variances and the covariance in the last line can be all determined
from experimental results with individual addressing.

To estimate the total excitation number variance $\Delta\hat{N}_1^2$, 
we need to perform
simultaneous measurement of internal and phonon states.
Since phonon states are usually measured by mapping them
to internal states, such a simultaneous measurement is not 
straightforward.
We here estimated the lower and upper bounds of the total excitation
number variance
from $\Delta\hat{N}_{a,1}^2$ and $\Delta\hat{N}_{p,1}^2$
to show the existence of polaritonic superfluidity.
\begin{eqnarray}
 \Delta\hat{N}_{1}^2
&=&
\Delta\hat{N}_{a,1}^2+\Delta\hat{N}_{p,1}^2
+2(\langle\hat{N}_{a,1}\hat{N}_{p,1}\rangle
-\langle\hat{N}_{a,1}\rangle\langle\hat{N}_{p,1}\rangle) \nonumber \\
&=&
\Delta\hat{N}_{a,1}^2+\Delta\hat{N}_{p,1}^2
+2 \mbox{Cov}(\hat{N}_{a,1},\hat{N}_{p,1})
\label{eqn:DeltaNc12}
\end{eqnarray}


Since $\hat{N}_{a,1}$ and $\hat{N}_{p,1}$ are positive operators,
$\langle\hat{N}_{a,1}\hat{N}_{p,1}\rangle\ge0$ and therefore
$\mbox{Cov}(\hat{N}_{a,1},\hat{N}_{p,1})\ge
-\langle\hat{N}_{a,1}\rangle\langle\hat{N}_{p,1}\rangle$.
In addition, from the Cauchy-Schwartz inequality for covariances,
$-\sqrt{\Delta\hat{N}_{a,1}^2 \Delta\hat{N}_{p,1}^2} \le
\mbox{Cov}(\hat{N}_{a,1},\hat{N}_{p,1}) \le
\sqrt{\Delta\hat{N}_{a,1}^2 \Delta\hat{N}_{p,1}^2}$ 
is satisfied.
In the end, 
$\mbox{Cov}(\hat{N}_{a,1},\hat{N}_{p,1})$
should satisfy the following inequality:
\begin{equation}
-\langle\hat{N}_{a,1}\rangle\langle\hat{N}_{p,1}\rangle 
\le
\mbox{Cov}(\hat{N}_{a,1},\hat{N}_{p,1}) \le
\sqrt{\Delta\hat{N}_{a,1}^2 \Delta\hat{N}_{p,1}^2} 
\label{eqn:CovNa1Np1}
\end{equation}
Thus, by using 
equation (\ref{eqn:DeltaNc12}) and (\ref{eqn:CovNa1Np1}),
the lower and upper bounds of $\Delta\hat{N}_{1}^2$ 
can be estimated with
the expectation values and variances of
the atomic excitation number and phonon number.